\documentclass[a4paper, 11pt]{article}
\usepackage[T1]{fontenc} 
\usepackage[utf8]{inputenc}
\usepackage[english]{babel}

\usepackage[a4paper,
            left=1in,
            right=1in,
            top=1.2in,
            bottom=1.6in]{geometry}

\usepackage{authblk}
\usepackage{graphicx}%
\usepackage{multirow}%
\usepackage{amsmath,amssymb,amsfonts}%
\usepackage{amsthm}%
\usepackage{hyperref}
\usepackage{mathrsfs}%
\usepackage[title]{appendix}%
\usepackage{xcolor}%
\usepackage{textcomp}%
\usepackage{manyfoot}%
\usepackage{booktabs}%
\usepackage{algpseudocode}%
\usepackage{listings}%
\usepackage{siunitx}
\usepackage[export]{adjustbox}
\usepackage{bm}
\usepackage{csquotes}
\usepackage[ruled,linesnumbered]{algorithm2e}
\usepackage{adjustbox}
\usepackage{url}
\usepackage{verbatim}
\usepackage{array}
\usepackage[caption=false,font=normalsize,labelfont=sf,textfont=sf]{subfig}
\usepackage{textcomp}
\usepackage{stfloats}

\DeclareSIUnit\cluce{c}

\graphicspath{{images/}}
\allowdisplaybreaks

\newcommand{\tsup}[1]{\textsuperscript{#1}}

\usepackage[sorting=none,citestyle=numeric-comp, natbib]{biblatex}
\bibliography{biblio}

\title{Fast and accurate noise removal by curve fitting using orthogonal polynomials}

\author[1]{Andrea \textsc{Gallo Rosso}\thanks{\href{mailto:andrea.gallo.rosso@fysik.su.se}
{andrea.gallo.rosso@fysik.su.se}.}}
\affil[1]{Physics Department and Oskar Klein Centre,\protect\\
Stockholm University, Stockholm, Sweden}

\begin{document}

\maketitle

\begin{abstract}\noindent
Local polynomial smoothing is a widespread technique in data analysis, and Savitzky–Golay (SG) filters are one of its most well-known realizations. In real settings, the effectiveness of SG filtering depends critically on proper tuning of its parameters, constrained in turn by repeated polynomial fitting over large data windows and for varying polynomial degrees. Standard implementations based on monomial bases and Vandermonde matrix formulations are known to suffer from ill-conditioning and unfavorable scaling as the problem size increases.
In this work, we present a fast and numerically stable method for computing polynomial fitting and differentiation matrices by reformulating the problem in terms of discrete orthogonal (Chebyshev) polynomials. Exploiting their recursive structure and the intrinsic symmetry properties of the resulting matrices, we derive two algorithms designed to reduce computational overhead. Both methods significantly reduce memory usage and improve scalability with respect to the polynomial degree and window length.
A discussion of the performance demonstrates that the proposed algorithms achieve orders-of-magnitude improvements in numerical accuracy compared to standard matrix multiplication, while also providing potential gains in execution time for large-scale problems. These features make the approach particularly well suited for applications requiring repeated local polynomial fits, such as the optimization of SG filters in high-resolution spectral analyses, including axion dark matter searches and the ALPHA haloscope.
\end{abstract}

\section{Introduction}
Curve fitting is a fundamental and omnipresent technique in any scientific discipline facing data analysis. Even in basic practices, parametric functions are crucial for representing observed data, extract meaningful features, and facilitate subsequent inference or prediction \cite[e.g.][]{bruce2020, grus2021}. In many practical settings, data are contaminated by noise. Therefore, not only fitting functions can provide predictive models that summarize complex phenomena, but they can also act as filters, to remove noise from raw measurements and to reveal the underlying signal structure.

Local polynomial smoothing (or local polynomial regression) provides a powerful and flexible framework to meet this two-fold requirement. It is widely used in applications like time-series analysis and spectral estimation, where the goal is to isolate deterministic patterns from stochastic disturbances, and especially when the signal trend is too complex for a single global function \cite[e.g.][]{scharf1991, lyons1997, shumway2017}. Unlike global regression, local polynomial smoothing fits a separate, simple model at every point of interest by defining a local neighborhood (or bandwidth), possibly applying weights via a kernel function, and then performing a least squares fit. The resulting ``smoothed'' value corresponds to the evaluation of this local polynomial at the target point.

Several methods have refined this approach; most notably, the locally weighted scatterplot smoothing (LOWESS) and locally estimated scatterplot smoothing (LOESS), which employ iterative weighting to ensure robustness against outliers \cite{Cleveland1979, Cleveland1988}. These methods have been extended to complex tasks like seasonal-trend decomposition (STL) \cite{Cleveland1990}. Another particularly relevant class of local polynomial models is the digital smoothing polynomial (DISPO) \cite{Ziegler:DISPO} and in particular the Savitzky–Golay (SG) filter \cite{SavGol}, which performs local polynomial regression on equally spaced data to preserve higher-order moments of the signal while suppressing noise. The versatility of SG filters has led to their adoption across a wide variety of fields such as biomedical engineering and physiology \cite{Hargittai:2005,rivolo2014au,Deepshikha:2016,agarwal2017,nishida2017,rivolo2017ac,sameni2017o},
analytical chemistry and molecular spectroscopy \cite{SavGol,Torabi2001,Schneider2003,Zimm2013,meng2014m,Agustika:2022},
environmental science and remote sensing \cite{Jonsson2004,Koluguri2017,Yang2017,Sibiya:2026},
industrial systems and energy infrastructure
\cite{Harnden1994,Kennedy2015,Suescin2016,Kher:2018,Seo2018},
and information and communication sciences
\cite{Koluguri2017,krishnan2013}.

This work is motivated by the application of SG filters to axion \cite{Peccei:1977hh, Peccei:1977ur, Weinberg:1977ma, Wilczek:1977pj, Kim:1979if, Shifman:1979if, Dine:1981rt, Zhitnitsky:1980tq} dark-matter searches \cite{Irastorza:2018dyq, Billard:2021uyg, Irastorza:2021tdu, Semertzidis:2021rxs, Adams:2855525, AxioBook2022} in general and to the ALPHA experiment \cite{ALPHA:2022rxj} in particular. It makes use of a resonant microwave cavity, commonly known as axion haloscope \cite{Sikivie:1983, Sikivie:1985yu}, to trigger the resonant conversion of axions into a detectable electromagnetic signal.
One of the primary challenges faced by haloscopes is the identification of an extremely weak, narrow-band signal, hidden within high-resolution power spectral densities. In this context, SG filtering is used to remove the complex, frequency-dependent background of the receiver system without artifacts that could mimic or obscure a real axion signal \cite[e.g.][]{Foster:2017hbq,Palken:2020wgs,GalloRosso:2022mhx}. SG filtering is valued for being analytical, numerically efficient, and predictable, while not requiring any background modeling. However, a careful implementation is needed to minimize some of its known issues such as numerical artifacts, biases, and signal underestimation \cite{Schmid:2022, Yi:2023ekw}.

The effectiveness of these filtering techniques is critically dependent on the selection of the polynomial degree and the window length. Together, they dictate the trade-off between noise suppression and signal distortion \cite{Sadeghi:2020, RamKrishnan}. Finding the optimal configuration is a non-trivial task. Many approaches can be found in the literature, ranging from manual trial-and-error approaches \cite{Rahman:2019}, to automated methods like grid searches \cite{Agustika:2022} or noise-residual matching \cite{Vivo:2006}. Statistical frameworks based on Stein’s \cite{Stein81}Unbiased Risk Estimator (SURE) have also been developed \cite{Seifzadeh:2011,RamKrishnan}.

However, these optimization strategies are characterized by a significant computational burden, where a smoothing operation is turned into a heavy, iterative search problem. For instance, we mention the need for iterative numerical solvers like Newton’s method to find optimal smoothing parameters \cite{Seifzadeh:2011}, the requirement for repeated evaluations across a range of candidate window lengths \cite{Zimm2013, Vivo:2006}, and the demand for pointwise adaptive filtering where the window is tuned for every single data point \cite{RamKrishnan}. Such procedures are frequently described as ``time-consuming'' \cite{Zimm2013}, especially if the fitting function and/or its first derivative needs to be evaluated across all the points in the windows. As a result, the computational cost can become prohibitively expensive for the high-resolution, large-scale datasets typical of modern axion haloscope experiments.

Traditional computational implementations of these fits often rely on power series or Vandermonde matrix formulations \parencite[e.g.][]{Guest,Schafer2011,NumRec2,NumRec3}. However, these become numerically burdensome as the polynomial order or the number of data points increases. Such approaches are prone to numerical ill-conditioning, where small perturbations in the data are amplified, leading to unstable estimates and undermining the reliability of the model.

A long-known and well-established alternative to the Vandermonde formulation in curve fitting is the use of orthogonal polynomials \cite[e.g.][]{Shohat,Szego,Guest,Cadwell1961S,Peck1962,Hayes1969A,Fan1996,Brinker} and specifically the so-called Chebyshev polynomials, named after Chebyshev and his work \cite{Chebyshev1853,Tchebychef}. In the context of repeated evaluations, a fitting process based on orthogonal polynomials can benefit from their recursive properties, allowing for the recursive addition of higher-order terms without recomputing the entire model. However, and as discussed in \cite{Brinker}, the evaluation of discrete Chebyshev polynomials via standard recurrence relations is prone to significant inaccuracies, particularly at high polynomial orders or for large data windows. These findings highlight that cumulative errors can lead to substantial deviations from the expected normalization, where the deviation can even exceed the norm itself as the interval size increases.

The goal of this work is to present a method that achieves both high numerical accuracy and improves computational efficiency. We present a recursive algorithm for the calculation of the polynomial fitting matrix that exploits its inherent bisymmetric properties to reduce memory usage and processing time. Our approach provides a fast and accurate solution for curve fitting in large-scale datasets, such as the high-resolution power spectra encountered in ALPHA. The presented algorithms are publicly available \cite{gitHubRepo}.

The paper is organized as follows. In Section~\ref{sec:theo}, we introduce the theoretical framework underlying curve fitting with orthogonal polynomials. Section~\ref{sec:NImpl} is devoted to the numerical implementation of the proposed method, where algorithmic details and practical considerations are discussed. The performance of the approach is then assessed in Section~\ref{sec:perf} through a series of numerical experiments designed to evaluate accuracy, robustness, and computational efficiency. Finally, Section~\ref{sec:conc} summarizes the main results of this work and outlines possible directions for future research.

\section{Curve fitting and orthogonal polynomials}
\label{sec:theo}
In this section, we provide a review of the basic 
concepts and formalism about curve fitting 
(Section \ref{sec:meth}), orthogonal
polynomials (Section \ref{sec:fit}), and their
numerical properties when applied to polynomial
fitting which will be used in Section
\ref{sec:NImpl}.

\subsection{Method and formalism}
\label{sec:meth}

Let us consider a set of $N$ points $\bm{x}$,
\begin{equation}
    \bm{x} = (x_0,\,x_1,\dots,x_{N-1})^T,
\end{equation}
and a set of $N$ data points $\bm{y}$,
\begin{equation}\label{eq:defY}
    \bm{y} = (y_0,\,y_1,\dots,y_{N-1})^T.
\end{equation}
The latter can be thought of as the outcome of a
measurement at different conditions represented by
each one of the $x_i$'s. 

Given the pairs $(x_i,\,y_i)$, we can define
a polynomial fitting function $f_n(x)$ as the
polynomial of degree $n$ that minimizes the
mean-squared error for the group of input samples:
\begin{equation}\label{eq:defEps}
    \mathcal{E}=\sum_{i=1}^N
    \left(f_n(x_i) - y_i\right)^2.
\end{equation}
Incidentally, we have assumed that the data points
are weighted equally. This assumption will hold
from now on.

Once the terms of the problems have been set,
the fitting function is univocally determined by
its set of coefficients $b_{j;n}$. That is,
\begin{equation}
    \label{eq:FitFun}
    f_n(x) = \sum_{j = 0}^n b_{j;n}\, x^j.
\end{equation}
The notation $b_{j;n}$ highlights that 
the coefficients depend both on the $j$-th power 
of $x$ and the degree $n$ of the fitting
polynomial.

Given expression \eqref{eq:FitFun}, we can write
equation \eqref{eq:defEps} as
\begin{equation}\label{eq:defEps2}
    \mathcal{E} = \sum_{i=1}^N\left(\sum_{j=0}^n
    b_{j;n}\,x_i^j - y_i\right)^2,
\end{equation}
where the only unknowns are precisely the 
$b_{j;n}$. A straightforward minimization process
yields the set of $n+1$ equations in $n+1$
unknowns that constitute the so-called normal
equations for the coefficients $b_{n;j}$:
\begin{equation}\label{eq:NEqBnj}
    \sum_{j=0}^n b_{n;j} \sum_{i=1}^{N} 
    \,x_i^j\,x_i^k =\sum_{i = 1}^{N} 
    y_i\,x_i^k \quad (k = 0,\dots,n).
\end{equation}
A well-established \parencite[e.g.][]{Guest,Schafer2011,
NumRec2,NumRec3} approach for addressing this problem is to express Equation \eqref{eq:NEqBnj} in matrix
form:
\begin{align}
    \label{eq:normal}
    \left(\bm{V}^T\cdot\bm{V}\right)\cdot \bm{b}
    &= \bm{V}^T\cdot\bm{y},\\
    \bm{b}&=\left(\bm{V}^T\cdot\bm{V}\right)^{-1}
    \cdot \bm{V}^T \cdot\bm{y}
    \label{eq:normal2}
\end{align}
where $\bm{V}$ is known as Vandermonde's matrix:
\begin{equation}
    \label{eq:defA}
    \left[\bm{V}\right]_{i\ell} = x_i^\ell.
\end{equation}
Thanks to expression \eqref{eq:normal2}, we can 
now express equation \eqref{eq:FitFun} as a
matrix product:
\begin{equation}
    f_n(x) = \bm{p}(x)\cdot\bm{b}
    \quad \text{with}\quad
    \bm{p}(x) = (1,x,\dots, x^n).
\end{equation}

As it will turn out to be useful in the following,
we can also define the vector $\bm{f}_n$
composed by the fitting function $f_n(x)$
evaluated at the points $x_i$:
\begin{equation}
    \bm{f}_n=(f_n(x_0),\,f_n(x_1),\dots,
    \,f_n(x_{N-1})),
\end{equation}
In this case, the vector of powers for each $x_i$ 
is precisely a row of the Vandermonde's matrix.
Thus, as it is straightforward to see
\begin{equation}
    \label{eq:defAmat}
    \bm{f}_n=\bm{V}\cdot\bm{b}=\bm{V}\cdot\left(
    \bm{V}^T\cdot\bm{V}\right)^{-1}\cdot \bm{V}^T
    \cdot\bm{y}\equiv \bm{A}_n\cdot\bm{y}.
\end{equation}

Thanks to the linear nature of the problem, we
have now a function that takes the vector of
measurements and automatically performs a
polynomial fit, returning the value of the
fitting function at the measured points.

In the following, we focus on the problem of
computing matrix $\bm{A}_n$. As definition 
\eqref{eq:defAmat} shows, a direct evaluation of 
the matrix product is not an optimal approach.
In fact, exponential growth \eqref{eq:defA} can
lead to accuracy losses, especially for large 
$n$ or $N$. These problems will be discussed in
Section \ref{sec:perf}. Moreover, and with respect
to the use of memory, the matrix $\bm{A}_n$ has
several symmetry properties (see Section
\ref{sec:AProp}). This means that the storage of
a full $N\times N$ matrix is not required---two,
in fact, as multiplication \eqref{eq:defAmat}
shows: $(\bm{V}^T\cdot\bm{V})^{-1}$ and $\bm{V}$. 

\subsection{Orthogonal polynomials}
\label{sec:fit}

In this section, we present an overview and some
basic properties of orthogonal polynomials, which
will be useful for the calculation of $\bm{A}_n$
and therefore for the optimization of the problem
of curve fitting. We refer to Refs.\ \cite{Guest,
Shohat} for notation and theoretical background.

In general, given a set of $x_i$'s, the set of
orthogonal polynomials of degree $n$ is defined by 
the relation 
\begin{equation}
    \label{eq:orto}
    \sum_{i = 1}^{N} p_n(x_i)\,p_m(x_i)
    = \delta_{nm} H_n^2,
\end{equation}
where 
$\delta_{nm}$ is the Kronecker delta and $H_n$
is the norm of the polynomial of degree $n$:
\begin{equation}\label{eq:HnFirst}
    H_n = \sum_{i = 1}^{N} p_n^2(x_i).
\end{equation}
We stress that, in general, $H_n \neq 1$.

The explicit expansion of $p_n(x)$ in powers of
$x$ will be written as
\begin{equation}
    \label{eq:Tjsum}
    p_n(x) = \sum_{j = 0}^n{\beta_{n;j}\,x^j}.
\end{equation}
Equation \eqref{eq:Tjsum} can also be  
inverted, leading to the expansion of $x$ as a 
series of $p_j(x)$:
\begin{equation}
    \label{eq:xnaj}
    x^n = \sum_{j = 0}^n \alpha_{n;j}\,p_j(x).
\end{equation}
An expression for $\alpha_{n;j}$ can be recovered
by combining relations \eqref{eq:xnaj} and
\eqref{eq:orto}:
\begin{equation}
    \sum_{i = 1}^{N}\left(x_i^n-\sum_{j = 0}^{n-1}
    \alpha_{n;j}\,p_j(x_i)\right)p_\ell(x_i)
    =\delta_{\ell n} \alpha_{n;n} H_n^2,
\end{equation}
which gives:
\begin{equation}
    \label{eq:alphNJ}
    \alpha_{n;j}=\frac{1}{H_n^2}\sum_{i = 1}^{N}
    x_i^n\,p_j(x_i).
\end{equation}

Thanks to relations
(\ref{eq:Tjsum}--\ref{eq:alphNJ})
we can express the normal equations  
\eqref{eq:NEqBnj} in terms of the orthogonal
polynomials $p_n(x)$. As a starting point, we
express the fitting function $f_n(x)$ in
\eqref{eq:FitFun} as
\begin{equation}
    \label{eq:FitFun2}
    f_n(x) = \sum_{j = 0}^n a_{j}\,p_j(x).
\end{equation}
It is worth noting that, unlike the $b_{n;j}$'s,
the coefficients $a_j$ do not depend on the
specific degree of the fitting function. In fact,
$f_{n+1}(x)$ can be computed by simply adding the
$(n+1)$\tsup{th} term to the series, while all the
$b_{n;j}$'s would need to be substituted by new
$b_{n+1;j}$'s.

A comparison between definitions \eqref{eq:FitFun}
and \eqref{eq:FitFun2} leads to:
\begin{equation}
    f_n(x) = \sum_{j=0}^n b_{n;j}\sum_{\ell=0}^j
    \alpha_{j;\ell}\,p_\ell(x)=\sum_
    {\ell=0}^n \left(\sum_{j=\ell}^n b_{n;j}\,
    \alpha_{j;\ell}\right)p_\ell(x)\equiv
    \sum_{\ell=0}^n a_\ell\,p_\ell(x).
    \label{eq:defFit2}
\end{equation}
Now, we can apply relations \eqref{eq:defFit2} and
\eqref{eq:alphNJ} to the l.h.s.\ of 
\eqref{eq:NEqBnj}, and relations \eqref{eq:xnaj}
and  \eqref{eq:alphNJ} to its r.h.s. This gives
\begin{equation}
    \label{eq:defaj}
    a_j = \frac{1}{H_n^2} \sum_{i = 1}^{N}
    y_i\,p_j(x_i),
\end{equation}
which finally leads to
\begin{equation}
    \label{eq:SGPol}
    f_n(x)=\sum_{i=1}^N y_i\sum_{j=0}^n\frac{p_j
    (x_i)\,p_j(x)}{H_j}.
\end{equation}

We can easily recognize the matrix product \eqref{%
eq:defAmat} hiding in relation \eqref{eq:SGPol}
and, consequently, how orthogonal polynomials 
link to matrix $\bm{A}_n$. Fundamental properties 
of $\bm{A}_n$ will be highlighted in Section
\ref{sec:AProp}. Before that, the explicit
expression for polynomials $p_j(x)$ will now be
discussed.

\subsection{Chebyshev polynomials}
\label{sec:Cheb}

The computation of polynomials $p_j(x)$ in
equation \eqref{eq:SGPol} was originally performed
by Chebyshev \cite{Chebyshev1853,Tchebychef,Shohat}.
In addition to the assumption of equally-weighted
data ($w_i=1$) his method works under the 
assumption that the series of points $x_i$ are
equally spaced by the same amount $\Delta x$:
\begin{equation}\label{eq:MustDx}
    \bm{x} =
    (x_0,\,x_0+\Delta x,\dots,x_0+(N-1)\,\Delta x).
\end{equation}
We stick to this assumption. We also follow
Chebyshev's choice to rescale $\bm{x}$ so as to
take up integer values:
\begin{equation}\label{eq:xWin0Np}
    x_\ell = \ell\in [0,\,1,\,\dots,\,N).
\end{equation}
The reasons for this choice are many. First of all,
the value of orthogonal polynomials is independent
of the origin of their variable, as e.g.\ pointed 
out in \cite[Sec. 7.2.1]{Guest}. Moreover,
equation  \eqref{eq:defAmat} and the matrix
$\bm{A}_n$ are independent of the equally-weighted
and equally-spaced points of application $x_i$'s.
Indeed, the $x_i$'s play no role beyond indexing, 
while the construction depends solely on the values
$y_i$'s and the associated $f_i$'s. Naming the
$x_i$'s from $0$ to $N-1$ will automatically link
them to the indices of the matrix $\bm{A}_n$.

In our notation, Chebyshev polynomials will be
denoted as $q_n(x)$. Under assumptions 
\eqref{eq:MustDx} and \eqref{eq:xWin0Np}, such
polynomials can be expressed using Szeg\"{o}'s
notation \cite{Szego} and the $n^{\text{th}}$
forward difference $\Delta^n$:
\begin{equation}
    \label{eq:defqn}
    q_n(\ell) = n!\,\Delta^n\!\left[\binom{\ell}{n}
    \binom{n - N}{n}\right].
\end{equation}
Polynomials $q_n(\ell)$ are not normalized, and
their norm $H_n$ \eqref{eq:HnFirst} is given by
\begin{equation}
    \label{eq:defHn1}
	H_n = (n!)^2\binom{2n}{n}\binom{N+n}{2n+1} =
    \frac{N \prod_{j = 1}^n
    \left(N^2 - j^2\right)}{2n + 1}.
\end{equation}

The crucial property that characterizes every set
of orthogonal polynomials is the existence of a
recursive relation that gives the polynomial of
degree $(n+1)$ from the ones of degree $n$ and
$(n-1)$ \cite[e.g.][]{Nikiforov}. In the case of
Chebyshev polynomials, the relation reads:
\begin{equation}
    \label{eq:ricorN} (n+1)\,q_{n+1}(\ell) =
    \left(2\ell-N+1\right)(2n+1)\,q_n(\ell)-
    n(N^2 - n^2)q_{n-1}(\ell),
\end{equation}
with initial values given by:
\begin{equation}
    \label{eq:1stval}
    q_0(\ell)=1\quad\text{and}\quad
    q_1(\ell)=2\ell-N+1.
\end{equation}
A similar relation holds also for coefficients
$\beta_{n;j}$ in \eqref{eq:Tjsum} as well as the
norm $H_n$. The former satisfies:
\begin{equation}
    \label{eq:ricorb}
 \beta_{n;j}=\frac{2(2n -1)}{n}\beta_{n-1;\,j-1}
    -\,\frac{(N-1)(2n-1)}{n}\beta_{n-1;\,j}
    -\frac{n-1}{n} \left(N^2 - 
    (n-1)^2\right)\beta_{n-2;\,j},
\end{equation}
where $\beta_{n;\,j} = 0$ if $j > n$ or $j < 0 $.
The norm $H_n$, on the other hand, satisfies:
\begin{equation}
    \label{eq:defHn2}
    H_n = \frac{2n-1}{2n+1} (N^2 - n^2) H_{n-1}.
\end{equation}

Chebyshev polynomials satisfy another kind of
recursive relation, this time in $\ell$, for fixed
values of $n$ (see e.g.\ \cite{Brinker}):
\begin{equation}
    \label{eq:ricoK}
	q_n(\ell) =D_n\,q_n(\ell-1)+ E_n\,q_n(\ell-2),
\end{equation}
assuming $n<N$ and $\ell\in[0,\,N)$.
The coefficients $D_n$ and $E_n$ are given by:
\begin{align}
	\label{eq:defDn}
	D_n &= -\frac{n(n+1)+(2\ell -1)
    (\ell - N -1) + \ell}{\ell (N - \ell)};\\
	\label{eq:defEn}
	E_n &=\frac{(\ell-1)(\ell-N-1)}{\ell(N-\ell)}.
\end{align}

Among the useful properties of Chebyshev
polynomials that will be useful in the following, we
mention their symmetry in $\ell\in[0,\,N)$
\cite{Nikiforov}. That is,
\begin{equation}
    \label{eq:syqsy}
	q_n(\ell) = (-1)^n\,q_n(N - 1 - \ell).
\end{equation}

In principle, what we have discussed so far would
be enough to determine the polynomials in equation
\eqref{eq:SGPol} and their coefficients 
$\beta_{n;j}$ in \eqref{eq:Tjsum}. Unfortunately,
however, their definition would remain implicit.
An explicit expression does exist, but we have to
resort to the so-called factorial notation borrowed
from \cite{Guest}. Let us write:
\begin{equation}
    \label{eq:qnFact}
    q_n(\ell) =\sum_{j=0}^n\beta_{(jn)}\ell^{(j)},
\end{equation}
where
\begin{equation}
    \label{eq:factorx}
    \ell^{(j)} \equiv\prod_{k= 0}^{j-1} (\ell- k) =
    \frac{\ell!}{(\ell - j)!}
\end{equation}
and
\begin{equation}
    \label{eq:benj}
    \beta_{(jn)} = (-1)^{n-j}
    \binom{n+j}{n}\binom{n}{j}
    \frac{(N -1)^{(n)}}{(N -1)^{(j)}}.
\end{equation}

Interestingly, coefficients $\beta_{(jn)}$ are
still related by a recursive relation. In fact, it
even takes a simpler form than \eqref{eq:ricorb}:
\begin{align}
    \label{eq:rbfat}
    \beta_{(j(n+1))} &= -\left(N-n -1\right)\,
    \frac{n+1+j}{n+1-j}\,\beta_{(jn)};\\
    \label{eq:rbfat2}
    \beta_{((j+1)n)} &= -\frac{n - j}{(j+1)^2}\,
    \frac{n+j+1}{N - j -1} \,\beta_{(jn)}.
\end{align}

We will make good use of these two relations in
Section \ref{sec:NImpl}, where we finally
tackle the problem of the computation of matrix
$\bm A_n$.

\subsection[Useful properties]{Useful properties of $\bm{A}_n$ and its derivative}
\label{sec:AProp}

At the core of our computation is the matrix
\begin{equation}
\label{eq:defAil}
    [\bm{A}_n]_{i\ell} =
    \sum_{j = 0}^n A^j_{i\ell}
    =\sum_{j = 0}^n\frac{q_j(i)\,q_j(\ell)}{H_j}.
\end{equation}
We remind that, in our notation, the row index
$i$ and the column index $\ell$ range from 0 to
$N-1$, where $N$ is the number of fitted points
and $n$ the degree of the fitting polynomial.

The first thing that emerges from definition
\eqref{eq:defAil} is the fact that matrix $\bm{A}
_n$ is bisymmetric. That is, and thanks to
relation \eqref{eq:syqsy}, $\bm{A}_n$ is symmetric
about both of its main diagonals:
\begin{subequations}\label{eq:diagRelA}
\begin{align}
    [\bm{A}_n]_{N-1-i,N-1-\ell}
    &=[\bm{A}_n]_{N-1-i,N-1-\ell} =\\
    &=[\bm{A}_n]_{\ell,i} =\label{eq:sub:diag}
    [\bm{A}_n]_{N-1-\ell,N-1-i},
\end{align}
\end{subequations}
where \eqref{eq:sub:diag} does not apply if
$i=\ell$ or $i+\ell= N-1$.
This means that the algorithm presented in Section 
\ref{sec:NImpl} will not have to deal with the
computation of $N\times N$ entries, but just with
a quarter of those. Moreover, a similar symmetry
holds for the two central axes:
\begin{align}
    \label{eq:mirsim}
    A_{i,N -1-\ell}^j &=(-1)^{j} A^j_{i\ell};\\
    \label{eq:mirror}
    [\bm{A}_n]_{i,N - 1 - \ell} &= \sum_{j = 0}^n
    (-1)^{j} A^j_{i\ell}.
\end{align}
Hence, while calculating $\bm{A}_n$ through 
summation \eqref{eq:defAil}, it will suffice to
know the terms $A^j_{i\ell}$ ($j=0,\dots,n$) for
only half of a quarter of the entries---see Figure
\ref{fig:mtAv0} and \ref{fig:mtA} for a graphical
representation.

Finally, the rows and columns of $\bm A_n$ sum up
to 1, thanks to the relation \eqref{eq:HnFirst},
at the core of the definition of orthogonal
polynomials. From a numerical perspective, dealing
with $\bm A_n$ as a whole helps preventing
overflows and loss of precision, in contrast with
going through $q_n(\ell)$  \eqref{eq:defqn} and
$H_n$ \eqref{eq:defHn1}  separately, or through
the exponential terms in Vandermonde's matrix
\eqref{eq:defA}.

As a side note, this means that $\bm{A}_n$ is
idempotent, too. That is, $\bm{A}_n\cdot\bm{A}
_n=\bm{A}_n$. This is a straightforward conclusion
from the fact that fitting the fitted values
returns the fitted values themselves; that is,
$\bm{A}_n\cdot\bm{f}_n=\bm{f}_n$.


Relation \eqref{eq:defAil} also allows for the
definition of the first derivative of the
polynomial fit \eqref{eq:FitFun}, evaluated at the
fitting points:
\begin{equation}
    \label{eq:defBil} 
    [\bm{f}_n']_i=\frac{1}{\Delta x}\sum_{\ell=0}
    ^{N-1} [\bm{B}_n]_{i\ell}[\bm{y}]_{\ell}
    =
    \sum_{\ell=0}^{N-1}
    \frac{y_\ell}{\Delta x}
    \sum_{j = 0}^n\frac{q_j'(i)\,q_j(\ell)}{H_j},
\end{equation}
where $\Delta x$ is the distance between the
data points. Note that, in our notation, we
preserve the subscript $n$ in $\bm{B}_n$, even if
the fitting polynomial has one lesser degree.

Matrix $\bm{B}_n$ is no more symmetrical. However,
it is anti-centrosymmetrical, as a differentiation
of relation \eqref{eq:syqsy} can easily show.
That is,
\begin{equation}\label{eq:symB}
    [\bm B_n]_{N-1-i,N-1-\ell}=-[\bm B_n]_{i\ell}.
\end{equation}
The same property holds for each of the terms
$B_{i\ell}^j$ in the sum \eqref{eq:defBil}.
Moreover, it can be shown that
\begin{subequations}\label{eq:Bijsy}
\begin{align}
    B^j_{N-1-i,\ell}&=(-1)^{j+1}B_{i\ell}^j;\\
    B^j_{i,N-1-\ell}&=(-1)^{j\hphantom{+1}}B_{i\ell}^j.
\end{align}
\end{subequations}

\section{Numerical implementation}
\label{sec:NImpl}

In this section, we present the algorithms to
compute $\bm{A}_n$ and $\bm{B}_n$. The computation
of $\bm{A}_n$ will follow two different approaches,
the first one being characterized by a better
numerical accuracy (Section \ref{sec:prec}), and
the other one by a faster execution (Section
\ref{sec:cmbf}). A systematic discussion about the
performances is presented in Section
\ref{sec:perf}.

\subsection[Numerical computation]{Numerical computation of $\bm{A}_n$}
\label{sec:prec}

\begin{figure}[t]
    \centering
    \includegraphics[width=0.8\textwidth]{%
    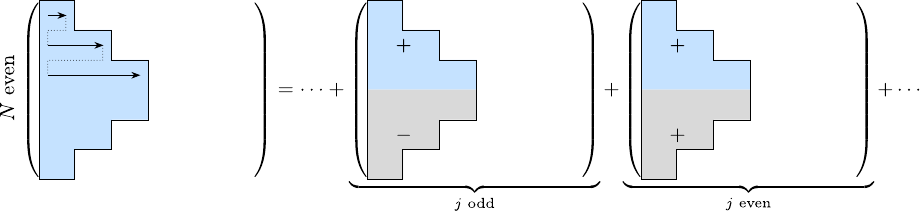}\\
    \includegraphics[width=0.8\textwidth]{%
    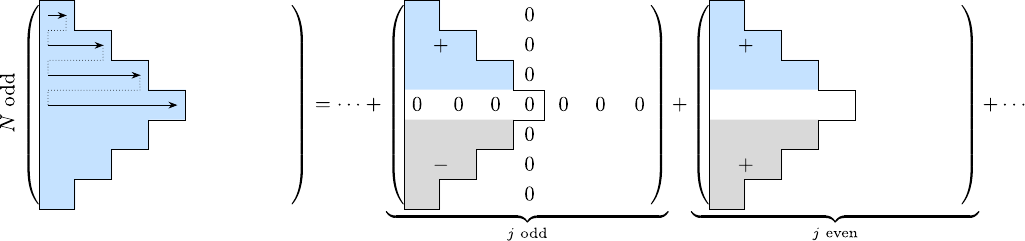}\\
	\caption{Graphical representation of the
    computation of the matrix $\bm{A}_n$ as
    described in Section \ref{sec:prec}. Because
    of symmetries \eqref{eq:diagRelA} only a
    quarter of it needs computation. And because
    of  \eqref{eq:mirror}, the sum for the entries
    in the lower part of the triangle can be
    recovered directly from the upper part}
	\label{fig:mtAv0}
\end{figure}

Based on definition \eqref{eq:defAil}, the
computation of each entry of $\bm{A}_n$ requires
the summation over the polynomial degree $j$;
i.e., over the set of values:
\begin{equation}\label{eq:defAvec}
    \vec{A}_{i\ell}=(A^0_{i\ell},\dots,
    \,A^n_{i\ell}).
\end{equation}
At each step, we take advantage of the recursive
relations to get $\vec{A}_{i\ell}$ from the values
that have already been computed.
The simplest way is to proceed by rows, as shown
in Figure \ref{fig:mtAv0}. The computation does
not need to span the whole set of columns because,
thanks to symmetries \eqref{eq:diagRelA} matrix
$\bm{A}_n$ is composed of 4 identical triangles.
In this case, we focus on the left one. Moreover,
values $[\bm A_n]_{i\ell}$ and $[\bm A_n]_{N-1-i,
\ell}$  are linked by relation \eqref{eq:mirror}.
Hence, the number of rows to be covered explicitly
is just\footnote{Incidentally, and for clarity
purposes, we point out that the code attached to
the present paper stores the result in a 
one-dimensional array of length $N_r\times(\lfloor
N/2\rfloor+1)$, where the index $k$ of the array
is linked to the row index $i$ and column index
$\ell$ by: $k=i+\ell(N-\ell)$.}: $N_r=\lfloor
N/2\rfloor+N\text{mod } 2$.

\SetKwComment{Comment}{$\triangleright$\ }{}
\DontPrintSemicolon
\SetKwFunction{addTo}{sum\_and\_update}
\begin{algorithm}[t]
\caption{Computation of $\bm{A}_n$
(Section \ref{sec:prec})}\label{alg:prc}
\KwData{$N$, $n < N$}
$i_{\max}
\gets \lfloor N/2\rfloor + (N\!\mod2) -1 $\;
$\bm{a}_0\gets\bm{a}_1\gets[0,\dots,0]$%
\Comment*[r]{$\vec A_{i,\ell-2}, \vec 
A_{i,\ell-1}$ in \eqref{eq:prgAil}}
\For{$i\leftarrow 0$ \KwTo $i_{\max}$}
{
\For{$\ell\leftarrow 0$ \KwTo $i$}
{
    \uIf{$\ell = 0$}
    {
    Compute $\bm{a}_1$ from eqs.\ (\ref{eq:A0i1},%
    ~\ref{eq:recAi0})\;
    }
    \uElseIf{$\ell =1$}{
    $\bm{a}_0\gets \bm{a}_1$\;
    Compute $\bm{a}_1$ from eq.\ \eqref{eq:Aj10}\;
    }
    \Else{
    Compute $\bm{a}_2$ from eq.\ 
    \eqref{eq:prgAil}\;
    $\bm{a}_1\gets \bm{a}_2$\;
    $\bm{a}_0\gets \bm{a}_1$\;
    }
    \addTo{$\bm A_n, \bm a_1$}\;
    }
}
\SetKwProg{myalg}{Subroutine}{}{}
\myalg{\addTo{$\bm A_n, \bm a_1$}}{
$\Sigma,\Sigma^\prime\gets0,0$
\Comment*[r]{Total of sum \eqref{eq:defAil} for
elements $(i,\ell)$ and $(i,N-\ell-1)$ of
$\bm A_n$}
$\sigma\gets+1$\Comment*[r]{Sign between $A^j_{i\ell}$ and $A^j_{i,N-\ell-1}$}
\For{$j\leftarrow 0$ \KwTo $n$}{
$\Sigma\gets\Sigma+\bm a_1[j]$\;
$\Sigma^\prime\gets\Sigma+\sigma *\bm a_1[j]$\;
$\sigma\gets-\sigma$
}
Write $\Sigma$ at element $(i,\ell)$ of $\bm 
A_n$\;
Write $\Sigma^\prime$ at element
$(N-i-1,\ell)$ of $\bm 
A_n$\;
}{}
\end{algorithm}

The computation relies on the recursive properties
of the orthogonal polynomials $q_n(x)$ embedded in
the definition of $\bm{A}_n$. In fact, for each
new row $i$, one just needs to obtain the first
two column entries---i.e.  $\vec A_{i0}$ and $\vec
A_{i1}$. Then, the recursive relation \eqref{%
eq:ricoK} will propagate them across the columns.
In turn, any $\vec A_{i\ell}$ can be obtained from 
the two values $A_{i\ell}^0$ and $A_{i\ell}^1$,
corresponding to a degree $j=0$ and $j=1$,
recursively propagated up to $n$ thanks to relation
\eqref{eq:ricorN}. The calculation of $A_{i\ell}
^0$ and $A_{i\ell}^1$ is straightforward and
derives directly from the explicit expression
\eqref{eq:1stval} for the Chebyshev polynomials
of degree $j=0$ and $j=1$, plugged into definition
\eqref{eq:defAil}:
\begin{equation}\label{eq:A0i1}
    A^0_{i\ell} =\frac{1}{N}\quad\text{and}\quad
    A^1_{i\ell} =\frac{3(N-1-2i)(N-1-2\ell)}
    {N(N^2-1)}.
\end{equation}

Then, the recursive relation for the subsequent
values of $\vec A_{i0}$ is derived by combining
relations \eqref{eq:ricorN} and \eqref{eq:defAil},
while applying relations (\ref{eq:defHn2},
~\ref{eq:rbfat}), and keeping in mind that
$q_n(0) = \beta_{(0n)}$. This gives:
\begin{equation}
    \label{eq:recAi0}
    A^j_{i0} =\frac{2j+1}{(N+j)\,j}\left((N-1-2i)
    A^{j-1}_{i0}\right. - \left.\frac{(j-1)(N-j+1)}{2j-3}
    A^{j-2}_{i0} \right).
\end{equation}

The computation of the second term in the row, $
\vec A_{i1}$, follows directly from $\vec A_{i0}$. 
We notice:
\begin{equation}
    q_n(1)=\beta_{(0n)}+\beta_{(1n)}=\left(
    1-\frac{n(n+1)}{N-1}\right)\beta_{(0n)},
\end{equation}
and therefore:
\begin{equation}
    \label{eq:Aj10}
    A^{j}_{i1}=\left(1 -
    \frac{j (j+1)}{N - 1}\right) A^{j}_{i0}.
\end{equation}

For any other index $\ell > 1$ we can use the
recursive relation in $\ell$, \eqref{eq:ricoK},
which remains valid even through definition
\eqref{eq:defAil}:
\begin{equation}
    \label{eq:prgAil}
    A^{j}_{i\ell} = D_n A^{j}_{i,\ell-1}
    + E_n A^{j}_{i,\ell-2}.
\end{equation}
We recall that $D_n$ and $E_n$ are defined in
(\ref{eq:defDn},~\ref{eq:defEn}) and depend on 
$\ell$. Algorithm \ref{alg:prc} summarizes the
whole procedure.

\begin{figure}[t]
    \centering
    \includegraphics[width=0.8\textwidth]{%
    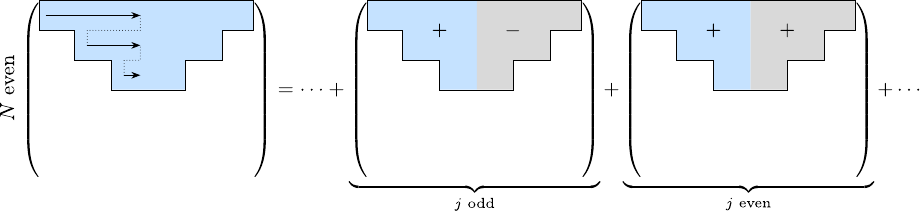}\\
    \includegraphics[width=0.8\textwidth]{%
    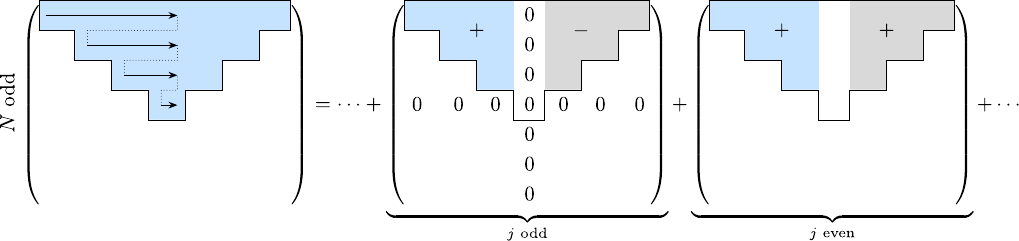}\\
	\caption{Graphical representation of the
    computation of the matrix $\bm{A}_n$ as
    described in Section \ref{sec:cmbf}. Because
    of symmetries \eqref{eq:diagRelA} only a
    quarter of it needs computation. And because
    of \eqref{eq:mirror}, the sum for the entries
    in the right part of the triangle can be
    recovered directly from the left part by
    a change of sign.}
	\label{fig:mtA}
\end{figure}

\subsection[Numerical computation An Buffer]{Numerical computation of $\bm{A_n}$. Buffer method}
\label{sec:cmbf}

\begin{figure}[t]
    \centering
    \includegraphics[width=\columnwidth]%
    {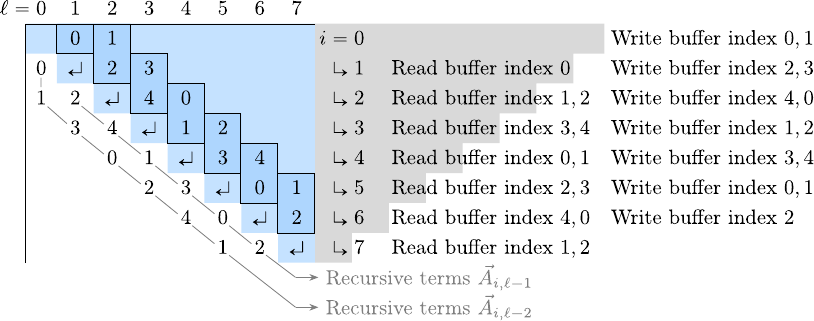}\\
	\caption{Magnification of the upper triangle
    of the right matrix in Figure \ref{fig:mtA}
    assuming $N=16$. Each new row in the
    computation process requires the recursive
    terms $\vec A_{i,\ell-2}$ and $\vec A_{i,\ell-
    1}$ to start the series. A circular buffer
    with indices from 0 to 4 ensures that the
    values can be read from the terms already
    computed; i.e., $\vec A_{\ell-2,i}$ and $\vec  
    A_{\ell-1,i}$.}
	\label{fig:Buff}
\end{figure}

Performing the calculation of $\vec{A}_{i0}$ and 
$\vec{A}_{i1}$ from scratch at the beginning of
every row gives Algorithm \ref{alg:prc} a certain
a certain degree ofnumerical stability. However,
one could improve the speed of Algorithm
\ref{alg:prc} by noting that this is not always
necessary.

Let us change perspective and focus on the top
triangle of $\bm A_n$, as shown by Figure 
\ref{fig:mtA}. In this case, the starting values
needed by each row are:
\begin{itemize}
    \item $\vec A_{00}$ and $\vec A_{01}$ for
    $i = 0$;
    \item $\vec A_{10}$ and $\vec A_{11}$ for
    $i = 1$;
    \item $\vec A_{i,i-2}$ and $\vec A_{i,i-1}$
    for $i > 1$.
\end{itemize}
Among them, only three need an ad-hoc calculation:
$\vec A_{00}$ and $\vec A_{01}$ for the first
row, and $\vec A_{11}$ for the second. All the
others are in fact computed at a certain point in
the previous steps, as the symmetries \eqref{%
eq:diagRelA} and Figure \ref{fig:Buff} make clear. 
A circular buffer of only 5 entries for the $\vec 
A_{i\ell}$ can store and read those values,
avoiding the need to pass through relation
\eqref{eq:recAi0} at every new iteration.

\begin{algorithm}[t]
\SetKwComment{Comment}{$\triangleright$\ }{}
\DontPrintSemicolon
\SetNoFillComment
\SetKwFunction{addTo}{sum\_and\_update}
\caption{Computation of $\bm{A}_n$}
\label{alg:fst}
\KwData{$N$, $n < N$}
$\bm{\mathrm{buffer}}\gets[\vec 0,\vec 0,\vec 0,
\vec 0,\vec 0]$\;
$\bm{a}_0\gets\bm{a}_1\gets[0,\dots,0]$%
\Comment*[r]{$\vec A_{i,\ell-2}, \vec 
A_{i,\ell-1}$ in \eqref{eq:prgAil}}
$i_{\max}
\gets \lfloor N/2\rfloor + (N\!\mod2) -1 $\;
\For{$i\leftarrow 0$ \KwTo $i_{\max}$}
{
\tcc{Initialize $\bm{a}_0,\bm{a}_1$}
\uIf{$i = 0$ (first row)}
{
    Compute $\bm{a}_0$ from eq.\ (\ref{eq:A0i1},%
    ~\ref{eq:recAi0})\;
    Compute $\bm{a}_1$ from eq.\ \eqref{eq:Aj10}\;
    Write $\bm{a}_1$ to \bm{\mathrm{buffer}}\;
    \addTo{$\bm A_n, \bm a_0$}\;
    \addTo{$\bm A_n, \bm a_1$}\;
}
\uElseIf{$i = 1$ (second row)}
{
    Read $\bm{a}_0$ from $\bm{\mathrm{buffer}}$\;
    Compute $\bm{a}_1$ from eq.\ \eqref{eq:Aj10}\;
    \addTo{$\bm A_n, \bm a_1$}\;
}
\uElseIf{$i > 1$}
{
    Read $\bm{a}_0,\bm{a}_1$ from
    $\bm{\mathrm{buffer}}$\;
}
\tcc{Propagate $\bm{a}_0,\bm{a}_1$ along the row}
$\ell_{\min}\gets\max(i, 2)$\;
$\ell_{\max}\gets
\lfloor N/2\rfloor+(N\!\!\mod2)-1$\;
\For{$\ell\leftarrow\ell_{\min}$ \KwTo
$\ell_{\max}$}{
    Compute $\bm{a}_2$ from eq.\ 
    \eqref{eq:prgAil}\;
    $\bm{a}_0,\bm{a}_1\gets\bm{a}_1,\bm{a}_2$\;
    \addTo{$\bm A_n, \bm a_1$}\;
    \uIf{$\ell=i+1$ \textbf{or} $\ell=i+2$}
    {
        Write $\bm{a}_1$ to \bm{\mathrm{buffer}}\;
    }
}
}
\SetKwProg{myalg}{Subroutine}{}{}
\myalg{\addTo{$\bm A_n, \bm a_1$}}{
$\Sigma\gets0$\Comment*[r]{Sum for element $(i,\ell)$}
$\Sigma^\prime\gets0$\Comment*[r]{Sum for element $(i,N-\ell-1)$}
$\sigma\gets+1$\Comment*[r]{Sign between $A^j_{i\ell}$ and $A^j_{i,N-\ell-1}$}
\For{$j\leftarrow 0$ \KwTo $n$}{
$\Sigma\gets\Sigma+\bm a_1[j]$\;
$\Sigma^\prime\gets\Sigma+\sigma *\bm a_1[j]$\;
$\sigma\gets-\sigma$
}
Write $\Sigma$ at element $(i,\ell)$ of $\bm 
A_n$\;
Write $\Sigma^\prime$ at element
$(i,N-\ell-1)$ of $\bm 
A_n$\;
}{}
\end{algorithm}

The same line of reasoning can also be applied to
increase the degree of $\bm A_n$ to $\bm{A}_{n+1}$,
without having to recompute all the previous
$n$ matrices $A^j_{i\ell}$ in sum \eqref{eq:defAil}.
In fact, it is enough to compute the three terms
$A^{n+1}_{00}$, $A^{n+1}_{01}$, and $A^{n+1}_{11}$
and then propagate those recursively, while adding
each $A^{n+1}_{i\ell}$ to the corresponding element
of $[\bm{A}_{n}]_{i\ell}$. Numerically, the 
problem is simplified by the fact that the
circular buffer is used to store and read just
$A_{\ell-2,i}^{n+1}$ and $A_{\ell-1,i}^{n+1}$,
rather than the whole set of values $\vec
A_{\ell-2,i}$ and $\vec A_{\ell-1,i}$.

\subsection[Numerical computation Bn]{Numerical computation of $\bm{B_n}$}
\label{sec:cmpB}

In the numerical computation of $\bm{B_n}$, we
cannot longer rely on bisymmetry. Therefore, the
calculation proceeds by rows, in the same fashion 
as Algorithm \ref{alg:prc}.
Once again,
the idea is to build $\vec{B}_{i0}$ and $\vec{B}_
{i1}$ for each new row, and use the recursive
properties of the Chebyshev polynomials to
increase the column index. Starting from $\vec{B}_
{i0}$, we notice:
\begin{equation}\label{eq:B0i1}
    B^0_{i\ell}=0;\qquad
    B^1_{i\ell}=-\frac{6(N-1-2\ell)}{N(N^2-1)}.
\end{equation}
Differentiation of equation \eqref{eq:recAi0}
provides the relation to obtain the values up to
$j = n$:
\begin{multline}
    \label{eq:recBi0}
    B^n_{i0}=\frac{2n+1}{(N+n)n}
    \left((N-1-2i)B^{n-1}_{i0} -
    2A^{n-1}_{i0}\vphantom{\frac{(n -1)(N -n +1)}
    {2n-3}}\right.\\
    \left.-\frac{(n -1)(N -n +1)}
    {2n-3}B^{n-2}_{i0}\right).
\end{multline}
In this case, however, and in contrast with
\eqref{eq:recAi0}, we note that the knowledge of
$A_{i0}^{n-1}$ is also required.

\begin{figure}[t]
    \centering
    \includegraphics[width=\columnwidth]%
    {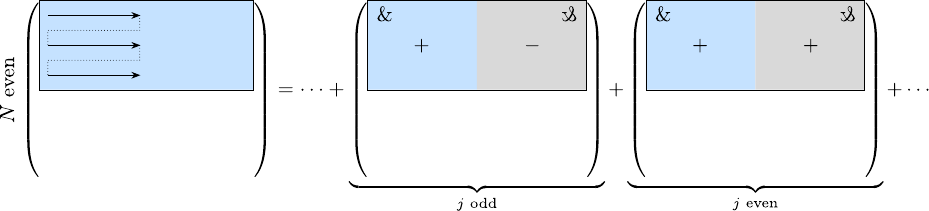}\\
    \includegraphics[width=\columnwidth]%
    {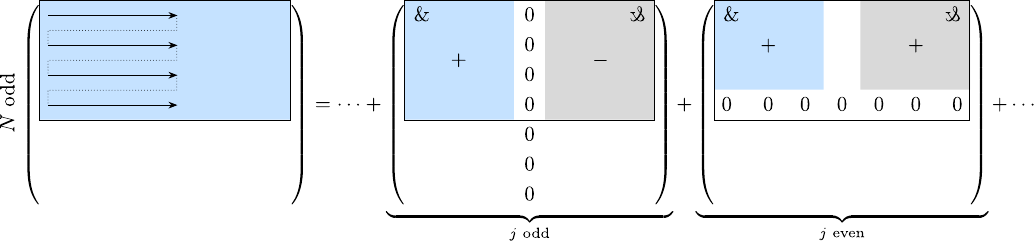}\\
	\caption{Graphical representation of the
    computation of the matrix $\bm{B}_n$. Because
    of symmetries \eqref{eq:symB} only the top
    half needs computation. And because of 
    \eqref{eq:Bijsy}, entries that are symmetrical 
    with respect to the central axis can be 
    recovered from the same $\vec{B}_{i\ell}$.}
	\label{fig:matBil}
\end{figure}

Once $\vec{B}_{i0}$ is known, $\vec{B}_{i1}$ is
recovered from a relation similar to \eqref{
eq:Aj10}:
\begin{equation}
    \label{eq:Bj10}
    B^{j}_{i1}=\left(1 -
    \frac{j (j+1)}{N - 1}\right) B^{j}_{i0}.
\end{equation}
Any further value along the row can be computed
from the same recursive relation in $\ell$, i.e.\
equation \eqref{eq:prgAil}. In fact, and from
definition \eqref{eq:defBil}, the differentiation
affects the polynomial corresponding to the index
$i$, while the one with index $\ell$ remains
unchanged.

As it happens for the computation of $\bm A_n$,
values $[\bm B_n]_{i\ell}$ and $[\bm B_n]_{N-1-i,
\ell}$  are linked by relation \eqref{eq:Bijsy}.
Hence, each step gets the two values symmetrical
with respect to the central axis, and the number
of rows to be covered by the loop is just $N_r =
\lfloor N/2\rfloor + N\text{mod } 2$ (see Figure
\ref{fig:matBil}).

\section{Performance}
\label{sec:perf}

\begin{figure}[t]
    \centering
    \includegraphics[width=\columnwidth]{%
    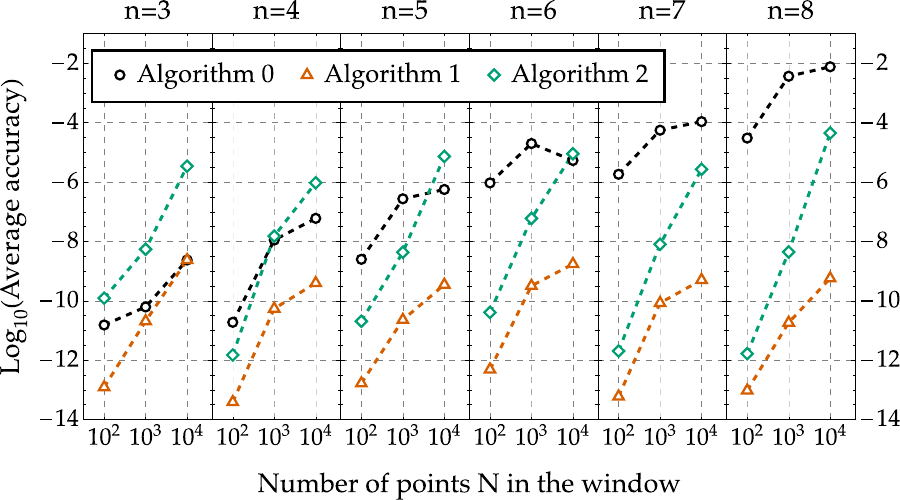}
	\caption{Comparison of the accuracy of 
    Algorithms 0, \ref{alg:prc}, and \ref{%
    alg:fst} in approximating matrix $\bm A_n$.
    The error distributions, quantified by the 
    standard deviation of the differences between 
    computed and true values (obtained from 
    Mathematica), are shown for varying polynomial 
    degrees $n$ and window lengths $N$.}
	\label{fig:errd}
\end{figure}

In this section, we evaluate the performances of
Algorithms \ref{alg:prc} (Section \ref{sec:prec})
and \ref{alg:fst} (Section \ref{sec:cmbf}). We
compare their execution time and accuracy against
direct matrix multiplication. For clarity, 
we refer to the direct matrix multiplication in
Eq.\ \eqref{eq:defAmat} as Algorithm 0. For the 
latter, we did not write our own routines, but
rather relied on the \texttt{TMatrixD} library of
ROOT \cite{ROOT}, which has the advantage of being 
already tested and optimized. In order to even out
contingent factors such as memory access and 
management, we use the homologous library 
\texttt{TArrayD} to handle the arrays in our
algorithm. The specific version of ROOT is
6.32.02, compiled with g\texttt{+}\texttt{+} (GCC)
11.4.1. The calculations were performed using
Almalinux 9 on a hypervisor ``Proxmox,'' with
processors Intel Xeon CPU E5-2650 v3 at 2.3 GHz.

The major difference emerging from the three
methods is the level of accuracy that they can 
reach, as shown in Figure \ref{fig:errd}. We
compared 2500 entries of $\bm A_n$ as given by
Algorithms 0, 1, and 2 with the true values
obtained from Mathematica with arbitrary
precision. The difference between the computed and
true values was then analyzed by examining the
distribution of these errors. To quantify the
accuracy of each algorithm, we report the standard
deviation of the error distributions, which serves
as an indicator of the variability in the errors
across all the elements of the matrix.

As a general trend, the accuracy worsens with
increasing number of points $N$ in the window.
However, the accuracy provided by Algorithms
\ref{alg:prc} and \ref{alg:fst} is more stable
across the values of the polynomial degree $n$,
while the direct matrix multiplication hits a
maximum of $\sim 1\%$ accuracy for $n=8$ and
$N=10^4$.
Algorithm \ref{alg:prc} is always more accurate
than Algorithm 0, although the latter is sometimes
more precise than the faster version (Algorithm
\ref{alg:fst}) for lower values of $n$.

\begin{figure}[t]
    \centering
    \includegraphics[width=0.7\textwidth]{%
    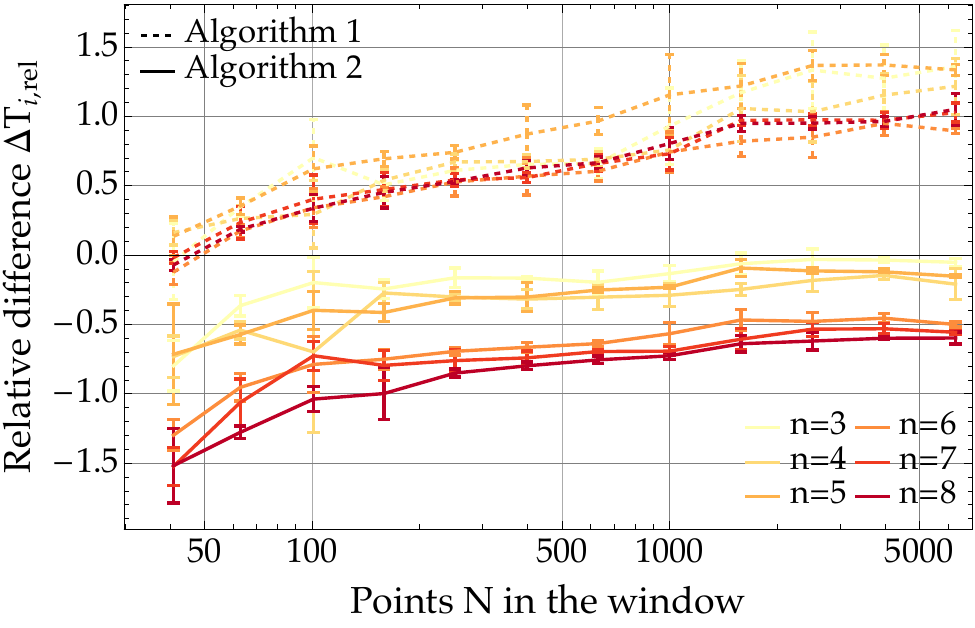}
	\caption{Relative difference \eqref{eq:Tdiff}
    in the time of
    computation of $\bm A_n$, between the full
    matrix multiplication \eqref{eq:defAmat} and
    Algorithms \ref{alg:prc} and \ref{alg:fst},
    presented in this work. The left
    panel shows the relative improvement as a
    function of points $N$ in the window, for
    different degrees $n$ of the polynomial fit,
    while the right panel shows the behavior as a
    function of the polynomial degree, for
    different window lengths.    
    }
	\label{fig:tim}
\end{figure}

As it is understandable, the trade-off for a better
accuracy impacts the time of execution. Figure 
\ref{fig:tim} shows the relative difference $\Delta
T_{i,\mathrm{rel}}$ in execution time $T_i$ for
Algorithms \ref{alg:prc} and \ref{alg:fst}, with
respect to Algorithm 0. We define it as
\begin{equation}
    \Delta T_{i,\mathrm{rel}} = \frac{T_i-T_0}
    {\min(T_0, T_i)},
    \quad\text{with i = 1, 2.}
    \label{eq:Tdiff}
\end{equation}

The result shows that Algorithm \ref{alg:fst} is 
always faster than Algorithm 0 and does not depend
(or does it very mildly) on the number $N$ of
points in the window. In fact, the dependency
rather concerns the polynomial degree $n$, with
improving performances as $n$ increases. Similar
trends are also shown by Algorithm \ref{alg:prc},
although it is generally slower than the standard
matrix multiplication. However, and going back to
the results of Figure \ref{fig:errd}, the slower
time is rewarded by an increase in accuracy by
orders of magnitude. For instance, given $n=8$ 
and $N$ between 100 and 1000, an increase in
computational time of $\Delta T_{1,
\mathrm{rel}}\approx 0.3-0.7$ is paid
off by an increase in accuracy of 8 orders of 
magnitude.

\section{Conclusion}
\label{sec:conc}

In this work we have addressed the problem of efficiently and accurately computing local polynomial smoothing via the calculation of fitting and differentiation matrices. While they can be expressed in closed form through standard least-squares formulations, direct implementations based on monomial bases and matrix multiplication are known to suffer from numerical instability and unfavorable scaling, particularly for large window lengths and moderate-to-high polynomial degrees.

By reformulating the problem in terms of discrete orthogonal (Chebyshev) polynomials, we have derived an explicit representation of the fitting matrix $\bm{A}_n$. Two numerical algorithms have been presented: a fully recursive approach optimized for numerical accuracy (Algorithm \ref{alg:prc}), and a buffer-based variant designed to minimize computational overhead (Algorithm \ref{alg:fst}). Both algorithms make use of the recursive properties of the orthogonal polynomials and the bisymmetric properties of $\bm{A}_n$ to reduce memory usage and the number of operations required for its evaluation. The same theoretical background also provides an efficient method to recursively increase the degree of the polynomial fit $n$.

A detailed performance study demonstrates that the proposed methods substantially improve numerical accuracy with respect to standard Vandermonde-based matrix multiplication, especially as the polynomial degree increases. In particular, the accuracy-optimized algorithm achieves improvements of several orders of magnitude in regimes relevant to high-resolution spectral analysis. At the same time, the buffer-based implementation consistently outperforms direct matrix multiplication in execution time, with only a mild dependence on the window size and increasingly favorable scaling for higher polynomial degrees.

These properties make the proposed approach well suited for large-scale applications such as axion dark matter searches. In this context, repeated polynomial fitting is required by algorithms that optimize the Savitzky–Golay parameters in order to extract narrow-band signals from slowly varying backgrounds, while avoiding distortions that could mimic or obscure a physical signal.

The present work assumes uniformly spaced data and equal weighting of samples, as is standard in Savitzky–Golay filtering. Extensions to weighted least-squares formulations or non-uniform sampling represent natural directions for future research.

\section*{Acknowledgments}
The author acknowledges support from the Knut and 
Alice Wallenberg Foundation and Olle Engkvists 
Foundation.

The author would like to thank Prof.\ Jan Conrad 
for helpful discussions and valuable insights that 
contributed to this work.

The author also acknowledge the support of the 
ALPHA Collaboration, whose collaborative 
environment was essential to this study.

\printbibliography

\end{document}